\documentstyle[epsf,rotate,psfig]{mn}

\begin{document}
\title[Primordial power spectrum]
{Features in the primordial power spectrum:
constraints from the CMB and the limitation of the 2dF and SDSS
redshift surveys to detect them
}
\author[\O. Elgar\o y et al.]
{\O ystein Elgar\o y$^1$,
 Mirt Gramann$^2$ and 
 Ofer Lahav$^1$ \\
 $^1$Institute of Astronomy, University of Cambridge, Madingley Road, 
Cambridge CB3 0HA, UK \\
 $^2$Tartu Observatory, T\~{o}ravere 61602, Estonia }
\maketitle
\begin{abstract}

We allow a more general (step-function) form of the 
primordial power spectrum than the usual featureless power-law 
Harrison--Zeldovich (with spectral index $n=1$) power spectrum, and fit it to 
the latest Cosmic Microwave Background  data sets.   
Although the best-fitting initial power spectrum can differ significantly from 
the power-law shape, and contains a dip at scales $k\sim 0.003\;h\,{\rm Mpc}^{-1}$,
we find that $\Omega_m \approx 0.24$, consistent  
with previous analyses that assume power-law  initial fluctuations.  
We also explore the feasibility of the early releases of the 2dF and SDSS galaxy 
redshifts surveys to see these features, and we find that even if features exist in the 
primordial power spectrum, they are washed out by the 
window functions of the redshift surveys on scales $k< 0.03\;h\,{\rm Mpc}^{-1}$.

\end{abstract}

\section{Introduction}

With the release of new Cosmic Microwave Background (CMB) 
data from DASI \cite{halv}, BOOMERANG \cite{nett} and 
MAXIMA \cite{lee}, and the 2dF redshift survey \cite{2dF}
nearing its completion, our ability to 
constrain cosmological models has improved significantly.  
The new CMB data removed the `baryon crisis' caused by the 
unexpectedly low amplitude of the second peak in the temperature 
power spectrum, and the standard model of the  universe now seems to 
be a flat Friedmann-Robertson-Walker universe, with 30\% matter 
($\sim$ 5\% baryons + $\sim$ 25\% cold dark matter (CDM)), 
70\% is dark energy, 
commonly parameterized by a cosmological constant, and with the current 
value of the Hubble parameter $H_0$ being around $70\;{\rm km}\;
{\rm sec}^{-1}\;{\rm Mpc}^{-1}$ (e.g. Efstathiou et al. 2001; 
Wang, Tegmark \& Zaldarriaga 2001).  

Some assumptions about the underlying cosmological model are necessary 
in order to extract these parameters from the data.  One common 
assumption is that the initial power spectrum of the density fluctuations 
is a featureless power law 
$P_{in}(k)\propto k^n$,  
where $k$ is the comoving wavenumber. 
The spectral index $n$ is found empirically 
to be close to the Harrison--Zeldovich 
value $n=1$.  This scale invariant primordial power spectrum is what 
typically comes out of models for inflation.  
However, there is no definite 
model for inflation, and some models predict features in 
$P_{in}(k)$.  For example, in supersymmetric double inflationary 
models, with two inflaton fields and two `trigger' fields, the 
power spectrum of density fluctuations is found to have a bump  
with superimposed oscillations on intermediate scales \cite{lesgo}.      
Features in $P_{in}(k)$ can also be produced in double inflation 
(Polarski \& Starobinsky 1992), and in one-field inflation with 
a feature in the inflaton potential (Starobinsky 1992).   
The fluctuation spectrum may be sensitive to physics at length 
scales below the Planck length \cite{brand,martin},   
and attempts have been made at  extracting mass fluctuation 
spectra from models inspired by string theory 
(Khoury et al. 2001; Kempf 2001;
 Kempf \& Niemeyer 2001;
Easther, Greene \& Shiu 2001).  In Kempf \& Niemeyer (2001) and 
Easther et al. (2001)  the primordial power spectrum was found to be of the 
Harrison--Zeldovich form, with a normalization depending on 
a short distance cutoff.  More realistic models will probably 
give rise to a $k$-dependent imprint \cite{east}.   It has been 
pointed out that models of this type  may run into problems in the 
form of an excessive creation of Planck energy particles at the 
present era (Starobinsky 2001), but it should at any rate be clear that  
the theoretical motivation for investigating more general forms 
for the initial power spectrum of density fluctuations is substantial.

On the observational side, several claims of indications for features  
in $P_{in}(k)$ have been made (Broadhurst et al. 
1990; Griffiths, Silk \& Zaroubi 2001; 
Atrio-Barandela et al. 2001; Barriga et al. 2001;
Hannestad, Hansen \& Villante 2001; Einasto et al. 
1999; Gramann \& H\"{u}tsi 2001; 
Silberman et al. 2001).   
Griffiths et al. \shortcite{griff} and Hannestad et al. \shortcite{hann}  
found that the CMB data favour a 
bump-like feature in the power spectrum at a scale 
$k\sim 0.004\;h\,{\rm Mpc}^{-1}$ ($h$ is the dimensionless 
Hubble parameter: $H_0 = 100h\;{\rm km}\,{\rm sec}^{-1}\,{\rm Mpc}^{-1}$).  
Barriga et al. \shortcite{barriga} introduced a step-like feature 
in the range $k\sim$ 0.06--0.6$\;h\,{\rm Mpc}^{-1}$ and found 
that this spectral break gave a good fit to both the CMB data 
and the data from the APM survey (Maddox et al. 1990).  
Atrio-Barandela et al. \shortcite{atrio} investigated the 
temperature power spectrum in CDM models with a matter 
power spectrum $P_m(k)$ at redshifts $z\sim 10^3$ of the 
form $P_m(k)\sim k^{-1.9}$ for $k > 0.05 \;h\,{\rm Mpc}^{-1}$.   
This form was derived by Einasto et al. 
\shortcite{einasto} by analysing observed power spectra of 
galaxies and clusters of galaxies. Gramann \& H\"{u}tsi \shortcite{gh} 
studied the mass function of clusters of galaxies with this 
form of $P_{in}(k)$ and found that the predicted number density 
of clusters was smaller than the observed one.  However, these 
authors found that they could get a good fit to the mass function 
with a $P_{in}(k)$ having a dip-like feature 
at $k\sim 0.1\;h\,{\rm Mpc}^{-1}$, and that this $P_{in}(k)$ also 
was consistent with data from other cosmological probes like 
peculiar velocities and CMB.  

One of the reasons for considering 
alternatives to a scale-invariant $P_{in}(k)$ was that the 
CMB data before April 2001 indicated that the amplitude of the second 
acoustic peak in the temperature power spectrum was   
low, resulting in  a baryon density $\Omega_b h^2 \sim 0.03$ 
outside the limits set by Big Bang Nucleosynthesis (BBN), 
which gives a 95 \% confidence interval $\Omega_b h^2 = 0.020 \pm 0.002$  
(Burles, Nollett \& Turner 2001).  
The new CMB data show  
a higher second peak, and the values for $\Omega_b h^2$ obtained with 
a power-law $P_{in}(k)$ are now consistent 
with standard BBN \cite{wang}.  

The motivation for the work 
presented in this paper is different: we want to check if the presently 
available CMB data allow for deviations from a scale-invariant $P_{in}(k)$, 
and if this is the case, whether these can be seen in the early releases 
of the 2dF and SDSS data.  
We will therefore consider more general shapes for $P_{in}(k)$ in our 
analysis.  

For simplicity, we will in our analysis vary only $\Omega_m$ and $P_{in}(k)$.  
We assume a flat Universe,  and consider 
parameters like $\Omega_b$ and $h$ to be well constrained by 
other cosmological probes.  The reader should note that these rather 
restrictive assumptions mean that the error bars we obtain for the 
estimated quantities will be much smaller than obtained in an analysis with 
more free parameters, like that of Wang et al. \shortcite{wang}.  Our aim in this 
paper is not to do the most general model fitting of the CMB+2dF, 
but to check two things: firstly, do the current CMB data allow 
features in $P_{in}(k)$ ? Secondly, can these features be seen in the 
early releases of the 2dF and SDSS data ?   

The outline of this paper is as follows: In Section 2 we introduce the 
models for $P_{in}(k)$ under consideration, and describe our procedure 
for fitting them to the CMB data.  In Section 3 we present the results 
of this procedure, and in Section 4 we discuss possible constraints from 
the 2dF and SDSS galaxy redshift surveys.  Finally, in Section 5 we 
summarize and discuss our results.  

\section{The primordial power spectrum and CMB anisotropies}

The power spectrum of fluctuations can be written as
\begin{equation}
P(k) = P_{in}(k) T^2(k), 
\label{eq:powdef1}
\end{equation}
where $T(k)$ is the transfer function (which modifies the initial 
power spectrum during the radiation dominated era). 
To investigate more general forms for $P_{in}(k)$, we let 
\begin{equation} 
P_{in}(k) = A k S(k), 
\label{eq:powdef2}
\end{equation}
where $A$ is a constant and $S(k)$ parameterizes the deviations from 
scale invariant initial fluctuations, and set up the models as follows.  
We modified the publicly available {\small CMBFAST} code \cite{seljak}  
to include two alternatives for $S(k)$: 
\begin{itemize}
 \item a `sawtooth'-shape, with `teeth' equally spaced in $\ln(k)$.  
 \item a set of `top-hat' steps, 
       equally spaced in $\ln(k)$ and with amplitudes $a_i$,$i=1,\ldots,N$ 
\end{itemize} 
To be specific, we defined the `sawtooth' spectrum following 
Wang \& Mathews \shortcite{wmat} as  
\begin{equation}
S(k)=\left\{\begin{array}{ll}
a_1  , & k \leq k_1 = k_{{\rm min}} \\
\frac{k_i-k}{k_i-k_{i-1}}a_{i-1}
+\frac{k-k_{i-1}}{k_i-k_{i-1}}a_i,& k_{i-1} < k < k_i \\ 
a_N, & k \geq k_{N} = k_{{\rm max}}, \end{array} \right . 
\label{eq:saw}
\end{equation}
where 
\begin{equation}
k_i = k_1 \left(\frac{k_{N}}{k_1}\right)^{(i-1)/(N-1)},\;i=2,\ldots,N-1, 
\label{eq:kdist}
\end{equation}
with $k_{{\rm min}} = 0.001\;h\,{\rm Mpc}^{-1}$, and 
$k_{{\rm max}}= 0.1 \; h \, {\rm Mpc}^{-1}$. 
Since the connection between the harmonic $\ell$ in the CMB power spectrum 
$C_\ell$ and $k$ is roughly $\ell \approx k d_A$ 
where for a flat universe the angular-diameter distance to the last scattering
surface is well approximated by (Vittorio \& Silk 1991):
\begin{equation}
d_A = \frac{2c}{H_0 \Omega_m^{0.4}},
\label{eq:angdist}
\end{equation}
where $c$ is the speed of light,   
the wave-number range $0.001 < k < 0.1\;h\,{\rm Mpc}^{-1}$ 
corresponds for $\Omega_m =0.3$ approximately to $ 10 < \ell < 1000$ 
which is nicely covered by the CMB data.

Alternatively, we let $P_{in}(k)$ be defined by a set of `top hat' steps, 
\begin{equation}
S(k)= \left\{\begin{array}{ll}
a_1 , & k \leq k_1 \\ 
a_i,  & k_{i-1} < k < k_i \\
a_N,  & k \geq k_{N-1}, \end{array} \right . 
\label{eq:step}
\end{equation}
with $k_i$ given by (\ref{eq:kdist}).  The comoving wavenumber $k$ 
is measured in units $h\;{\rm Mpc}^{-1}$, where $h$ is the dimensionless 
Hubble parameter.  

Both spectra are thus completely specified by 
$N$ and the values of $a_{1},\ldots,a_{N}$.  In our calculations 
we chose $N=4$.  We also let the matter density $\Omega_m$ be  
a free parameter, but the other parameters were kept fixed at 
$H_0 = 72\;{\rm km\;s}^{-1}{\rm Mpc}^{-1}$  \cite{hst}, 
$\Omega_b h^2 = 0.02$ \cite{BBN}, $N_\nu = 3.04$ (see e.g. 
Bowen et al. 2001), and we assumed a flat universe with no massive 
neutrinos and no reionization.      
We were therefore left with a five-dimensional parameter space to search 
for the best-fitting model.  This was done by minimizing the $\chi^2$ statistic  
given by 
\begin{equation}
\chi^2({\bf p}) = \sum_{ij}[\Delta {\cal T}_{i}^2-\Delta T_{i}^2({\bf p})]
(C^{-1})_{ij} [\Delta {\cal T}_{j}^2-\Delta T_{j}^2({\bf p})],
\label{eq:chisq}
\end{equation}
where ${\bf p}=(\Omega_m,a_1,a_2,a_3,a_4)$ is the parameter vector, 
$\Delta {\cal T}_i^2$ are the measured CMB fluctuations, 
\begin{equation}
\Delta T_i^2({\bf p}) = \frac{T_0^2}{2\pi}\sum_{\ell} W_{i\ell}
\ell(\ell + 1)C_\ell ({\bf p}), 
\label{eq:Ttheory}
\end{equation}
$C_\ell ({\bf p})$ is the COBE-normalized output from {\small CMBFAST}, 
$T_0=2.726\;{\rm K}$, $W_{i\ell}$ is the window 
function, and $C_{ij}$ is the covariance matrix of the observations. 
The observations, $\Delta {\cal T}_i$, $W_{i\ell}$, and $C_{ij}$ are taken from \cite{wang}. 
(The absence of the usual $1/\ell$-factor in Eq. (\ref{eq:Ttheory}) 
is a result of their definition of the window function).    
We computed the $\chi^2$ on a grid of $10^5$ models, 
and located the region of parameter space containing the global minimum.  
In this region we performed a more accurate search for 
the optimal parameters using the downhill simplex method \cite{numrec}.

\section{Results} 

\begin{figure} 
\begin{center}
{\centering
\mbox
{\psfig{figure=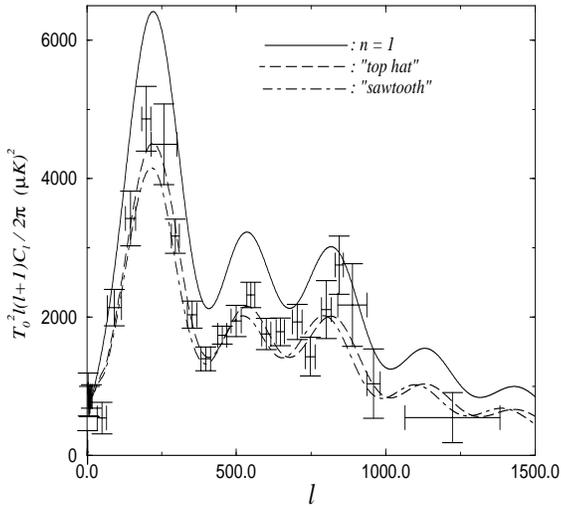,height=8cm,width=8cm}}
}
\caption{The best-fitting `sawtooth' and `top hat' models compared with the data.  
For comparison, we also plot a model where we have fixed $n=1$, $\Omega_m=0.24$, 
and the other parameters ($\Omega_b$, $h$ etc.) at the values given in the text, 
and with COBE normalization, which is higher than most of the 
new data points.}
\label{fig:fig1}
\end{center}
\end{figure}

\begin{figure}
\begin{center}
{\centering
 \mbox
 {\psfig{figure=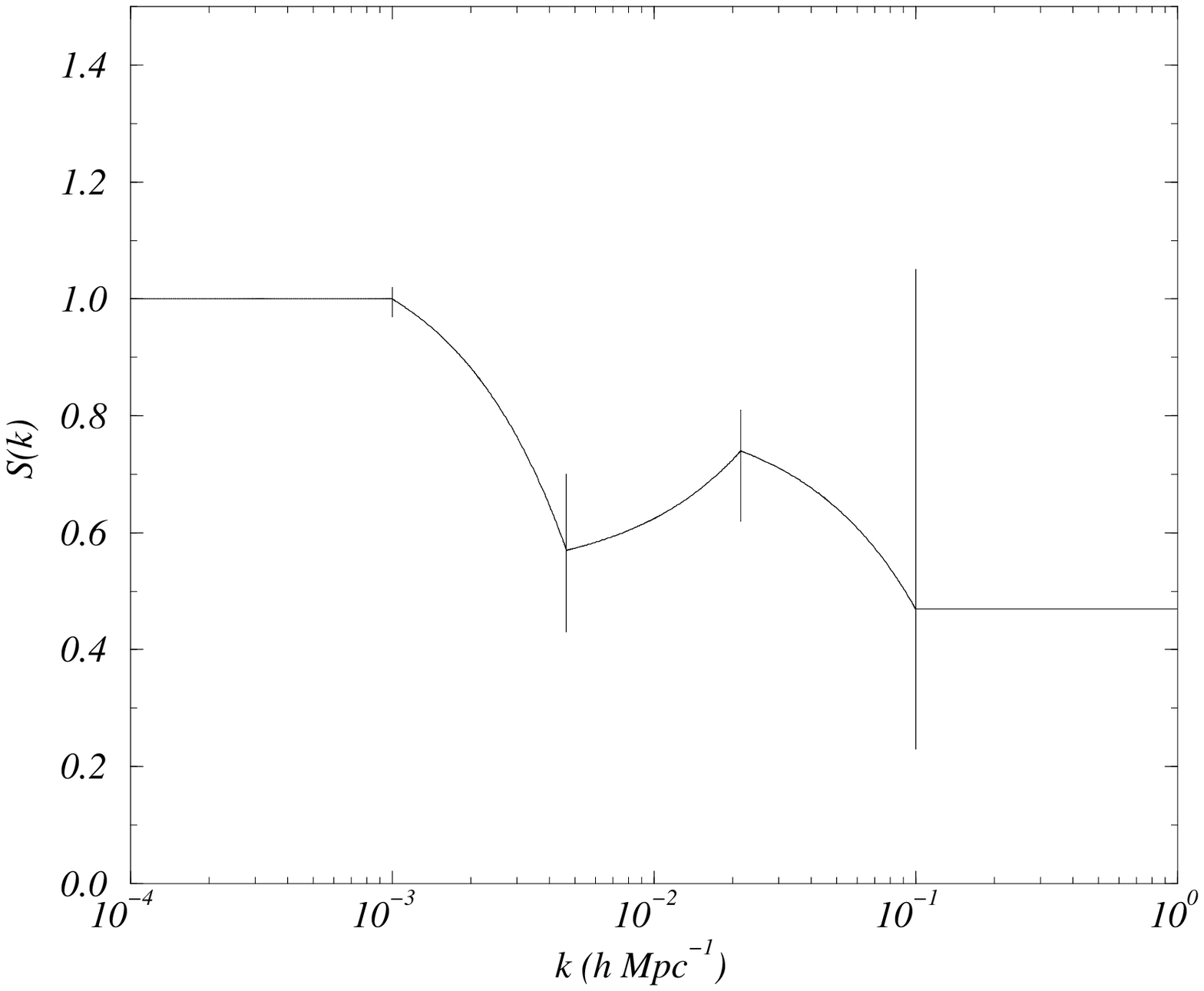,height=4cm,width=4cm}
 \psfig{figure=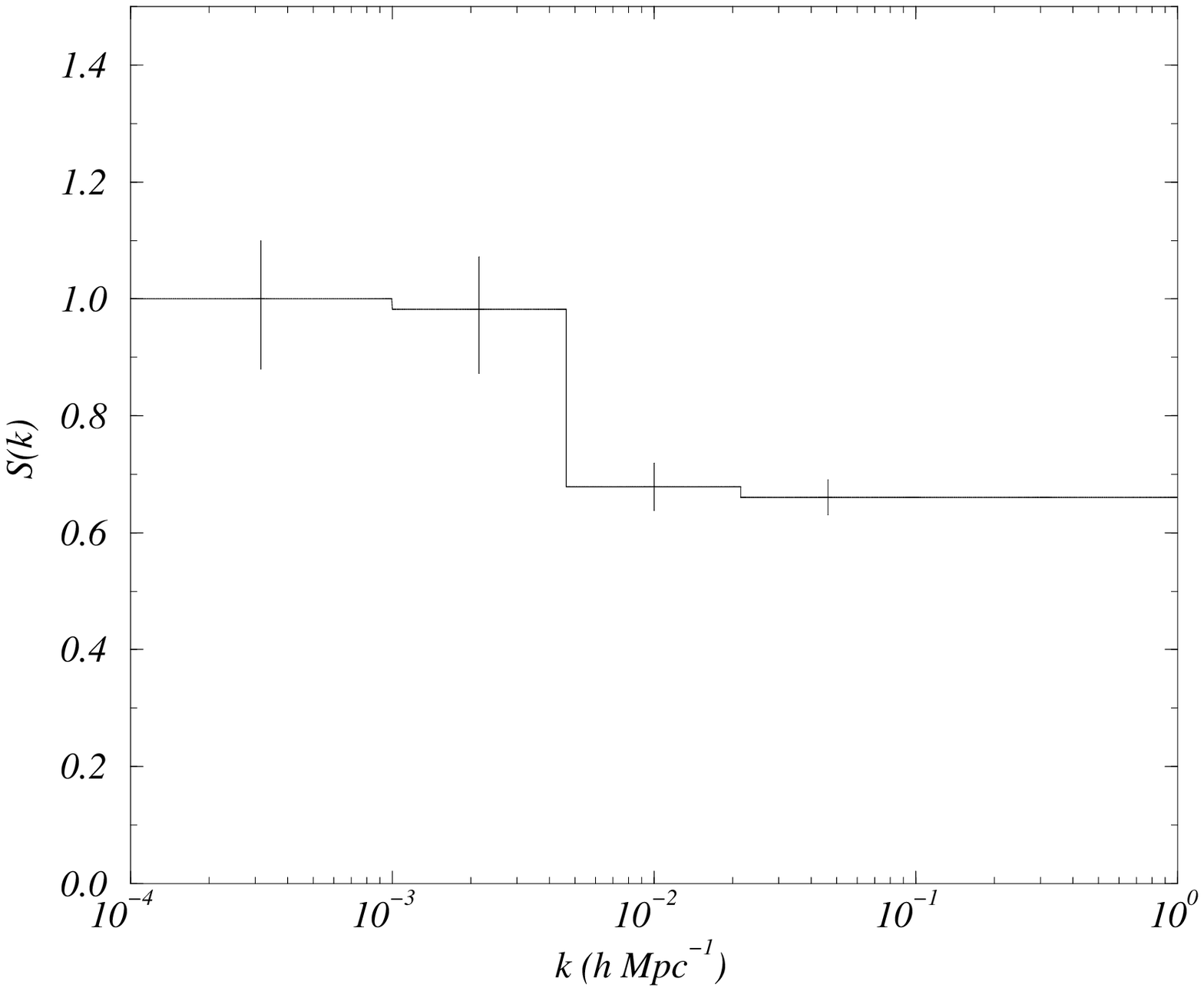,height=4cm,width=4cm}
 }
 \caption{The best-fitting `sawtooth' and `top hat' $S(k)$.  Vertical lines 
indicate the error bars on the estimated parameters.  The $k$ values spanned in 
this figure corresponds roughly to values of $\ell$ in the range 1--$10^4$. }
\label{fig:fig2}
}
\end{center}
\end{figure}

In Fig. \ref{fig:fig1} we show the best models for the two 
types of $P_{in}(k)$ we consider.  For comparison we have also 
plotted the power spectrum for a model with $P_{in}(k)=Ak$ (i.e. 
$S(k)\equiv 1$ and $\Omega_m = 0.24$).  
The best-fitting `top hat' and `sawtooth' models both have 
$\chi^2 \approx 32$ for the 24 data points.  
The high $\chi^2$ values are partly caused by the 
bandpowers centered at $\ell = 2$ and $\ell = 50$ in the 
compilation of Wang et al.  
Removing these points leads to lower values for the minimum $\chi^2$, but 
has no significant effect on our estimated values 
for $\Omega_m$ and the parameters of $P_{in}(k)$.  Note that the 
CMB only constrains the quantity $\Omega_m h^2$, but since we have  
fixed $h$ in our analysis, we obtain an estimate for $\Omega_m$ directly.    
In Fig. \ref{fig:fig2} we show the $S(k)$ which provide 
the best fit to the data, with $a_1$ scaled to 1 for clarity.  
The best-fitting parameters, mean values and 
confidence intervals can be found in Table 1.   
The confidence intervals for the parameters were obtained by the standard 
approach of constructing marginalized likelihoods for each parameter 
by integrating out the other parameters from the likelihood ${\cal L}\propto \exp(-\chi^2/2)$.
The tight confidence intervals, in particular on $\Omega_m$, reflect the restrictive 
assumptions we have made about other cosmological parameters. 
 
We see from Fig. \ref{fig:fig2} that both the best-fitting $S(k)$ have a dip 
at $k\sim 0.003\;h\,{\rm Mpc}^{-1}$.  
In fact, for the `top hat' $P_{in}(k)$ the data  favour 
$a_1 \approx a_2$ and $a_3 \approx a_4$.  This means that there are 
effectively only two parameters in this spectrum: the position of the 
dip and the relative sizes of the amplitudes.  We have checked this 
by repeating the analysis with more amplitudes in the spectrum, and 
found that the data in this case still favour a $P_{in}(k)$ with 
a dip at $k\sim 0.003 \;h\,{\rm Mpc}^{-1}$.     

\begin{figure}
\begin{center}
{\centering
 \mbox
 {\psfig{figure=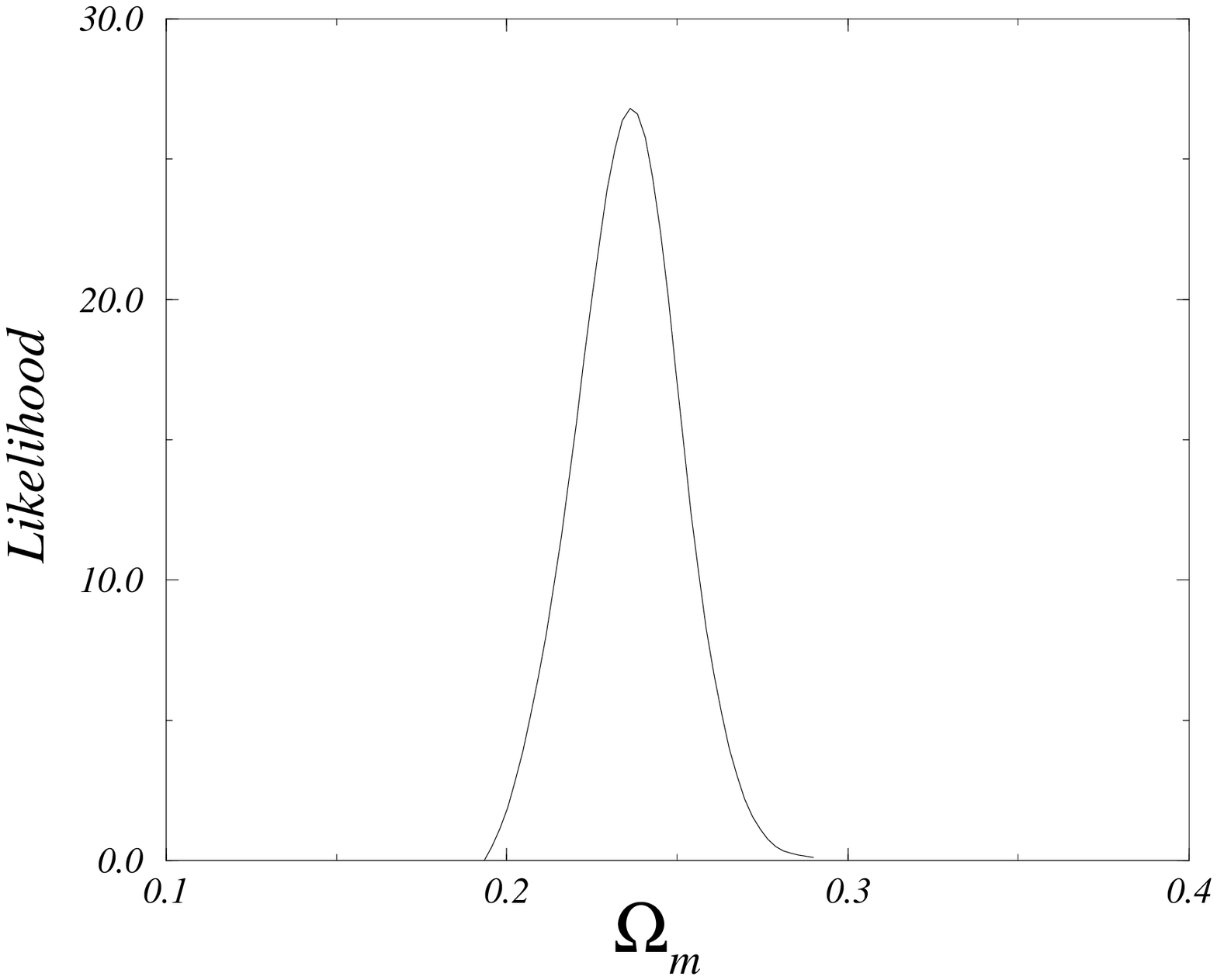,height=3cm,width=3cm}
 \psfig{figure=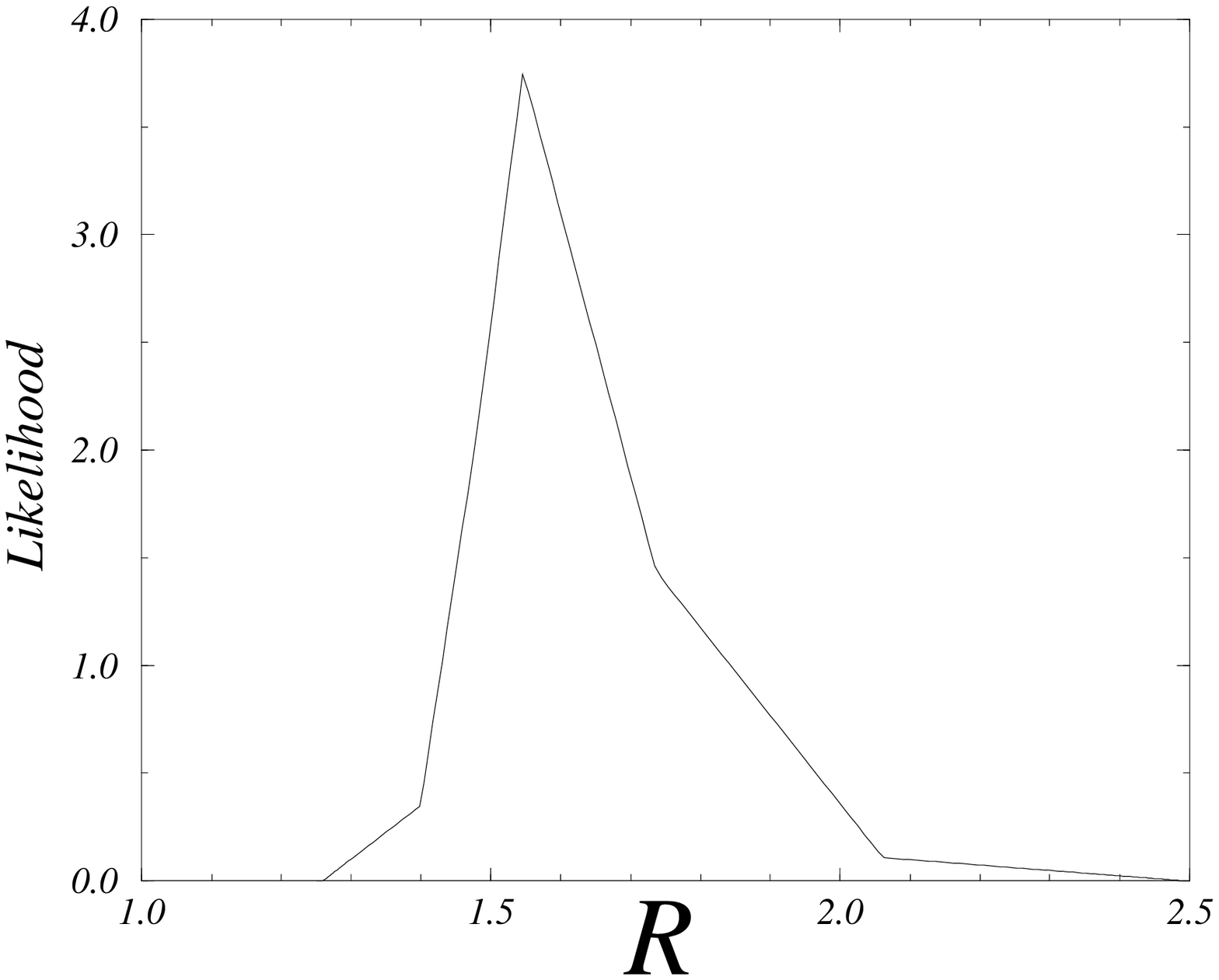,height=3cm,width=3cm}
 \psfig{figure=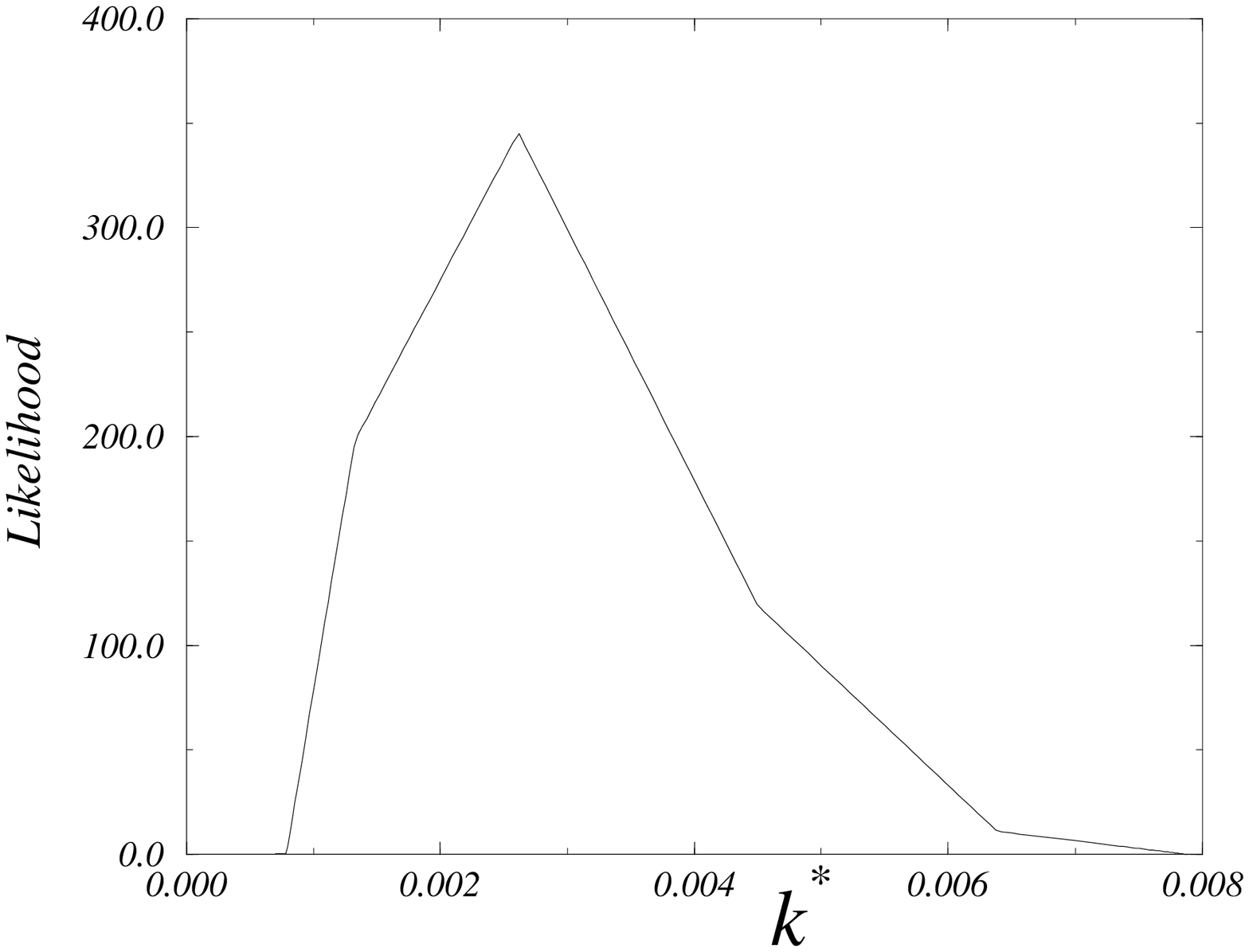,height=3cm,width=3cm}
 }
 \caption{Marginalized likelihood distributions for the 
 parameters in the `top hat' model.}
\label{fig:fig3}
}
\end{center}
\end{figure}

Thus, in this case we can reduce the parameter space to three dimensions: 
$\Omega_m$, the ratio ${\cal R}$ between the two amplitudes defining 
$P_{in}(k)$, 
and the position in $k$-space $k^*$ of the dip.   The best-fitting values and 
confidence limits are given in Table 2.  
For completeness, we also made a calculation for the 
`sawtooth' spectrum with this reduced set of parameters (i.e. keeping 
just one `tooth' in the spectrum, and using its amplitude and position 
in $k$-space as free parameters, see the tables).    
The marginalized likelihood distributions for $\Omega_m$, ${\cal R}$, and 
$k^*$  are shown in Fig. \ref{fig:fig3}.   We see that $\Omega_m$ is well 
constrained to a narrow range around 0.24, and the size of the dip ${\cal R}$ 
similarly constrained to be around $1.6$, consistent with the results for four steps.  
The scale $k^*$ at which the break occurs has a broader distribution.  In the 
calculation with four steps, this scale was at $\sim 4.6\cdot 10^{-3}\;h\,{\rm Mpc}^{-1}$, 
but from Fig. \ref{fig:fig3} and Table 2 we see that when we allow $k^*$ to vary, 
we can only constrain it to be in the range $\sim 0.001$--$0.005\;h\,{\rm Mpc}^{-1}$. 
\begin{table}
\begin{center}
\caption{Best-fitting values and 95 \% confidence intervals for the parameters of the two models.}
\begin{tabular}{@{}|l|l|l|l|l|@{}} 
\hline 
\multicolumn{1}{r}{  } &
\multicolumn{1}{r}{`Top hat'}&
\multicolumn{1}{r}{ } & 
\multicolumn{1}{r}{`Sawtooth'} & 
\multicolumn{1}{r}{ }\\
\multicolumn{1}{c}{ }               &
\multicolumn{1}{c}{Best fit} & 
\multicolumn{1}{c}{95\% c.i.}  &
\multicolumn{1}{c}{Best fit} &
\multicolumn{1}{c}{95\% c.i.}  \\ \hline 
$\Omega_m$    & $0.24$ & $0.24_{-0.02}^{+0.02}$  & $0.29$ & $0.29_{-0.02}^{+0.02}$   \\   \hline
 $a_1$        & $0.56$ & $0.59 _{-0.10}^{+0.12} $  & $1.0$ &$1.00_{-0.03}^{+0.02}$  \\ \hline 
 $a_2$        & $0.55$ & $0.59_{-0.09}^{+0.11}  $  & $0.57$ & $0.57_{-0.14}^{+0.13}$ \\   \hline 
 $a_3$        & $0.38$ & $0.40 _{-0.04}^{+0.04}$  & $0.74$ & $0.70_{-0.12}^{+0.07}$   \\  \hline  
 $a_4$        & $0.37$ & $0.38 _{-0.03}^{+0.03}$  & $0.47$ & $0.38_{-0.24}^{+0.58}$   \\   \hline 
\end{tabular}
\label{tab:table1}
\end{center}
\end{table} 
\begin{table}
\begin{center} 
\caption{95 \% confidence intervals for the parameters of the reduced 
models with two bins.  As in Table 1, the central values are the mean values. 
$k^*$ is given in units of $10^{-3}\;h\,{\rm Mpc}^{-1}$.} 
\begin{tabular}{|l|l|l|l|l|}
\hline 
\multicolumn{1}{r}{ } &      
\multicolumn{1}{r}{`Top hat'} &  
\multicolumn{1}{r}{ } & 
\multicolumn{1}{r}{`Sawtooth'} & 
\multicolumn{1}{r}{ } \\ 
\multicolumn{1}{c}{ } & 
\multicolumn{1}{c}{Best fit }&
\multicolumn{1}{c}{95\% c.i.}&
\multicolumn{1}{c}{Best fit}&
\multicolumn{1}{c}{95\% c.i.}\\ \hline
$\Omega_m$ &$0.24$&  $0.24_{-0.04}^{+0.02}$ &$0.24$& $0.23_{-0.03}^{+0.03}$ \\ 
${\cal R}$ &$1.55$&  $1.63_{-0.28}^{+0.21}$ &$1.74$& $1.77_{-0.34}^{+0.26}$ \\ 
$k^*$      &$2.62$& $2.9_{-2.0}^{+1.6}$ &$5.0$& $4.7_{-4.1}^{+3.0}$
 \\ \hline 
\end{tabular}
\label{tab:table2}
\end{center}
\end{table}

\section{Could 2\lowercase{d}F and SDSS detect features in the primordial 
power spectrum ?}

Combining data from different cosmological probes can in many cases lead to 
tighter constraints on the cosmological parameters, see e.g. Efstathiou et al. 
\shortcite{gpe} and Wang et al. \shortcite{wang}.  Since we saw in the previous 
section that the CMB does not rule out deviations from a scale-invariant 
$P_{in}(k)$, it is interesting to see if we can obtain further constraints 
from the matter power spectrum as estimated from the 2dF and SDSS galaxy 
redshift surveys. 

\begin{figure}
\begin{center}
{\centering
 \mbox
 {\psfig{figure=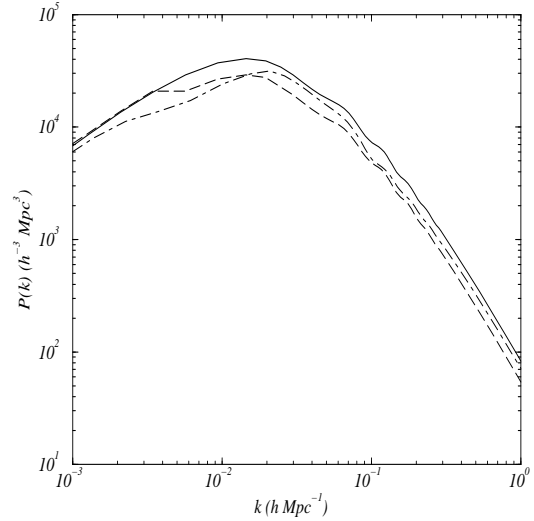,height=8cm,width=8cm}
 }
 \caption{Theoretical power spectra from CMBFAST.  The solid line 
corresponds to 
$S(k)\equiv 1$ ($\Omega_m=0.24$, $h=0.72$), 
the dashed line to the best-fitting `top hat' $S(k)$ model, and the dash-dotted 
line to the best-fitting `sawtooth' $S(k)$ model.}
\label{fig:fig4}
}
\end{center}
\end{figure}

Assuming a simple scale-independent biasing model
with  a bias parameter $b$, the galaxy power spectrum
(linear theory) is predicted to be 
\begin{equation}
P_{g}(k) = b^2 P_{m}(k)
= b^2 A k S(k) T^2 (k)
\label{eq:matterp}
\end{equation}
where $T(k)$ is the transfer function.  The recent analysis of 
2dF+CMB data by Lahav et al. (2001) found $b=1.0\pm 0.1$ on comoving 
scales of $0.02<k< 0.15\;h\,{\rm Mpc}^{-1}$.  
Using CMBFAST, we computed $P(k)$ 
for $S(k)\equiv 1$ and the best-fitting `top hat' and `sawtooth' $S(k)$, with 
the results shown in Fig. \ref{fig:fig4}.   
However, to compare with the galaxy power spectrum from 2dF (Percival et al. 2001), 
we must convolve $P_g(k)$ with the 2dF window function:  
\begin{equation}  
P_{\rm conv}(k) \propto \int P_{g}(|{\bf k}-{\bf q}|) \langle 
|W_q|^2\rangle d^3 q ,
\label{eq:2dFconv}
\end{equation}
where $\langle |W_q| ^2 \rangle $ is the spherical average of the the 2dF 
window function, approximately given by 
\begin{equation}
\langle |W_k| ^2 \rangle = \frac{1}{1+(k/a)^2 + (k/b)^4},
\label{eq:2dFwin}
\end{equation}
with $a=0.00342$, and $b = 0.00983$.  
As shown in Appendix A, the convolution integral can be rewritten as 
\begin{equation}
P_{\rm conv}(k) \propto \int_0 ^\infty K(k,k') P_{g}(k')dk',
\label{eq:convred}
\end{equation} 
where $K(k,k')$ is given by an integral over the 2dF window function (\ref{eq:2dFwin}) 
that can be evaluated analytically.   
As discussed in Appendix A, for $k\sim 0.1\;h\,{\rm Mpc}^{-1}$, 
$K(k,k')$ is almost a 
delta function $\delta(k'-k)$, so in this region $P_{\rm conv}(k) = P(k)$.  
For low $k$, $K(k,k')$ is a broad distribution, and the main contribution 
to the convolution integral comes from values of $k'$ larger than 
$\sim 0.01\;h\,{\rm Mpc}^{-1}$, so that $P_{\rm conv}(k)$ 
is nearly independent of $k$.  

\begin{figure}
\begin{center}
{\centering
 \mbox
 {\psfig{figure=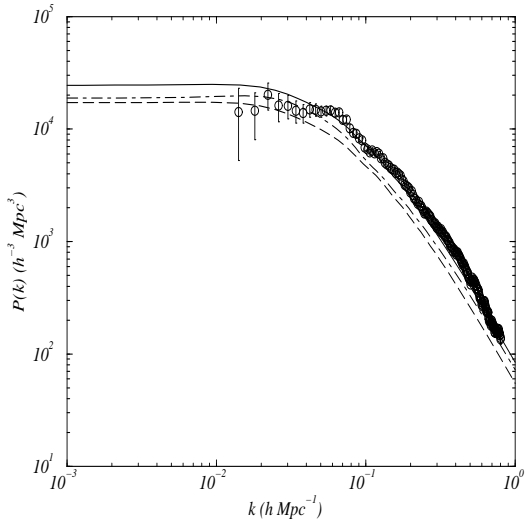,height=8cm,width=8cm}
 }
 \caption{Theoretical matter power spectra  
convolved with the 2dF window function and compared with the 2dF data.  
The solid lines corresponds to 
$S(k)\equiv 1$ ($\Omega_m=0.24$, $h=0.72$),
 the dashed line to the `top hat' $S(k)$, the dash-dotted 
line to the `sawtooth' $S(k)$, and the vertical bars are the data points from 
Percival et al. (2001). }
\label{fig:fig5}
}
\end{center}
\end{figure}

As a result, the features in $P(k)$ 
introduced by $S(k)$ are washed out by the convolution, as can be seen 
from Fig. \ref{fig:fig5}.  The results show that the present 
data from the 2dF survey cannot give us information
about the power spectrum at wavenumbers smaller than
about $0.03\;h\,{\rm Mpc}^{-1}$,
everything on larger scales is washed out.  
There is no relation 
between the wiggles visible in 2dF power spectrum, which is the result of 
observing a single realization of the true power spectrum convolved with 
the window function of the survey, and the features we introduced in $P_{in}(k)$.  
The wiggles in the 2dF power spectrum may be signatures of baryon oscillations, 
but may also be a  result of correlated noise.     
The scale of the observed wiggles is $\Delta k = 0.03\;\,{\rm Mpc}^{-1}$, which 
is the same scale over which power is correlated, so the wiggles could well 
be the result of correlated noise.  This tentative conclusion is supported by 
the recent analysis of the published sampled carried out 
by Tegmark, Hamilton \& Xu (2001).   

\begin{figure}
\begin{center}
{\centering
 \mbox
 {\psfig{figure=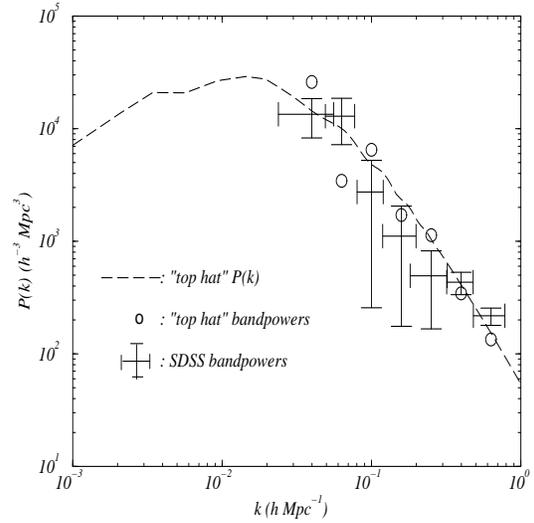,height=8cm,width=8cm}
 }
 \caption{Comparison of the `top hat' $P_g(k)$ with the   
SDSS power spectrum in the magnitude bin $r^*=21-22$.  Note that, in contrast 
with Fig. 5, it is the deconvolved power spectra that are shown in this 
figure.}
\label{fig:fig6}
}
\end{center}
\end{figure}

We also compared our calculated $P_g(k)$ with the recent estimate of the power 
spectrum from the Sloan Digital Sky Survey (SDSS), taken from 
Dodelson et al. \shortcite{dodelson}, where the SDSS data were analysed to obtain the 
deconvolved three-dimensional power spectrum.   
The SDSS results are given as a set of bandpowers in $k$-space, and thus 
the calculated power spectrum $P_g(k)$ must be transformed in a way analogous to 
Eq. (\ref{eq:Ttheory}); in the $i$th bin, the bandpower 
$P(k_i)$ is given by 
\begin{equation} 
P(k_i) = \sum_j W_{\rm SDSS}(i,j)P_g(k_j).  
\label{eq:pbin}
\end{equation}
where $k_i,k_j$ are the central values of the $k$ bins.  The window  
functions $W_{\rm SDSS}$ can be found in Dodelson et al. (2001).    
Our results are shown in Fig. \ref{fig:fig6}. 
Only results for the magnitude bin $r^* = 21-22$ 
are shown, but the same conclusion is true for the other three magnitude bins 
given in Dodelson et al. (2001).      
We see that the same conclusion applies to the SDSS spectrum as to the 
one from 2dF: at present there is no information on the scale where we find features in 
$P_{in}(k)$.  However, this conclusion only applies to the presently available data, as 
both 2dF and SDSS will, once completed, give information about 
larger scales than those probed by the data used in this paper.  

\section{Conclusions} 

We have analysed the new and updated CMB data,   
relaxing the usual assumption of power-law initial fluctuations.  
We found that the value of $\Omega_m$ we could extract 
was consistent with other analyses, e.g. $\Omega_m = 0.24 ^{+0.02}
_{-0.04}$ for the `top hat' $P_{in}(k)$ with varying position of the 
break.  We should point out that the small error bar on $\Omega_m$ found in 
our analysis is mainly a result of the restrictive priors we put 
on other quantities.  For example, allowing the Hubble constant to vary would have 
led to a significant increase in the uncertainty in $\Omega_m$ since the CMB power 
spectrum depends on $\Omega_m$ only through the physical matter density 
$\omega_m = \Omega_m h^2$.

We find that the present CMB data allow the initial power spectrum of 
the density fluctuations, $P_{in}(k)$ to have quite significant features, 
in particular a dip at a comoving wavenumber $k\sim 0.003\;h\,{\rm Mpc}^{-1}$, 
which corresponds to $\ell \sim 40$.  This dip can be understood 
as follows:  Increasing $\Omega_mh^2$ decreases the amplitudes of the 
peaks in the CMB power spectrum, 
and also shifts their positions to lower multipoles $\ell$.  
The position of the first peak is well determined by the data, and the 
models pay a higher price in terms of the $\chi^2$ for not fitting it 
than they do for not fitting the amplitude.    The position of the first 
peak in the compilation of CMB data we have used is well fitted by $\Omega_mh^2=0.12$ 
(Wang et al. 2001), which for $h=0.72$ gives $\Omega_m=0.24$.    
Fig. \ref{fig:fig1} shows that the CMB power spectrum of a model with this  
value of $\Omega_m h^2$, but a Harrison--Zeldovich $P_{in}(k)$, lies consistently 
above   the data points for all but the lowest values of $\ell$.  
The freedom in the `sawtooth' and `top hat' $P_{in}(k)$ 
allows the amplitude to be fitted by reducing $P_{in}(k)$ at comoving wavenumbers 
above $k\sim 0.003\;h\,{\rm Mpc}^{-1}$.

From the preceding discussion it also follows that the parameters of the 
`sawtooth' and `top hat' $P_{in}(k)$ are sensitive to calibration errors in the data points, 
and also that the choice of normalization for the models plays a role.  
We have normalized all our CMB power spectra to COBE, using the prescription given 
by Bunn \& White (1997).  Recent work by Lahav et al. (2001), 
Reiprich \& Boehringer (2001), Seljak (2001), and Viana, Nichol \& Liddle (2001), 
found that $\sigma_8$, the rms 
mass fluctuations in $8\;h^{-1}\,{\rm Mpc}$ spheres, is $\sim 20\%$ lower 
than predicted by the COBE normalization.  If we had used these results 
to normalize our CMB power spectra, the size of the dip we found 
in $S(k)$ would have been reduced.   
However, our goal here was to test the maximal features that can still 
be consistent with the CMB data.  Therefore the analysis with the 
constraint of COBE normalization can be viewed as an upper limit on the 
amplitude of the features.  

We also compared the early-release 2dF and SDSS galaxy power spectra to the theoretical 
predictions from our best-fitting $P_{in}(k)$, and we found that the observed power 
spectra are not sensitive to the features in $P_{in}(k)$ at comoving 
scales $k < 0.03\;h\,{\rm Mpc}^{-1}$.  There is no relation between the 
features we find in $P_{in}(k)$ and the wiggles observed in the 2dF 
power spectrum.  

Furthermore, we point out again that we have fixed all parameters 
except $\Omega_m$ and $P_{in}(k)$ in our analysis.  Combining 
extra degrees of freedom in $P_{in}(k)$ with a full-scale analysis of 
the CMB data, would lead to larger error bars on the cosmological 
parameters.    The more accurate measurements of the CMB fluctuations expected in  the 
near future, analysed jointly with  data sets from other cosmological probes,  
will hopefully allow us to put tighter constraints both on the 
primordial fluctuations and the parameters defining the geometry of the 
Universe. 

\section*{Acknowledgments}
We thank Will Percival for providing the 2dF data points, and   
Sarah Bridle, Andrew Hamilton, Raven Kaldare, Alexei Starobinsky, 
and Are Strandlie for useful comments.  MG has been supported 
by the ESF grant 3601.  
\O E is supported by a post-doctoral fellowship from The 
Research Council of Norway (NFR).

\appendix
\section{The convolution integral} 
The convolution of the power spectrum $P(k)$ with 
the window function is given by 
\begin{equation} 
\hat{P}_{\rm conv} ({\bf k}) \propto \int P_g ({\bf k}-{\bf q})
|W_{\bf k}({\bf q})|^2 d^3 q .
\label{eq:appeq1}
\end{equation}
Assuming $P$ is isotropic, and that only the spherical 
average of the final power spectrum is of interest, one can 
replace the window function with its spherical average, which for 
2dF can be approximated by 
\begin{equation}
\langle |W_k|^2\rangle = \frac{1}{1+\alpha k^2 +\beta k^4},
\label{eq:appeq2}
\end{equation} 
where $\alpha = 8.55\cdot 10^4$, $\beta = 1.071\cdot 10^8$.  
Choosing ${\bf k}$ as the $z$-axis in the integration, the convolution integral 
can be written as   
\begin{eqnarray} 
\hat{P}_{\rm conv} (k) &\propto& \int P(|{\bf k}-{\bf q}|) 
\langle |W_q|^2\rangle d^3 q \nonumber \\ 
& = & 2\pi \int_0 ^\infty dq q^2 \langle |W_q|^2\rangle \nonumber \\  
&\times & \int_{-1}^{+1}d(\cos\theta)P_g(\sqrt{k^2 + q^2 -2kq\cos\theta}),  
\label{eq:appeq3} 
\end{eqnarray}
where $\theta$ is the angle between ${\bf k}$ and ${\bf q}$.  
(Here and in the following we ignore the normalization constant. In 
practical calculations it is taken care of by dividing by $\int \langle 
|W_q|^2\rangle d^3 q$.)   
On substituting $k'^2 = k^2 + q^2 -2kq\cos\theta$, 
$d(\cos\theta) = -k'dk'/kq$, the integral becomes 
\begin{equation} 
\hat{P}_{\rm conv} (k)\propto \frac{2\pi}{k}\int_0^\infty dq q 
\langle |W_q|^2\rangle 
\int_{|k-q|}^{k+q}dk' k' P_g(k'). 
\label{eq:appeq4}
\end{equation} 
With the simple approximation Eq. (\ref{eq:appeq2}) to the 2dF window 
function, the convolution integral can be simplified further by 
changing the order of integration in Eq. (\ref{eq:appeq4}).  
We integrate over  
the region in the $(k',q)$-plane bounded by the lines $k'=k+q$, 
$k'=k-q,\;q\leq k$, and $k'=q-k,\;q>k$, so changing the order of 
integration is easy, and the result is 
\begin{eqnarray}
\hat{P}_{\rm conv} (k) &\propto& \frac{2\pi}{k}
\int_0^\infty dk' k' P_g(k') 
\int_{|k'-k|}^{k'+k}dq q \langle |W_q|^2\rangle \nonumber \\ 
&\equiv& \int_0^\infty K(k,k')P_g(k') dk' \label{eq:appeq5}  
\end{eqnarray} 
where  
\begin{equation}
K(k,k') \equiv \frac{2\pi k'}{k} \int_{|k'-k|}^{k'+k}dq q \langle |W_q|^2\rangle. 
\label{eq:appeq6}
\end{equation}
With the 2dF window function, one can obtain an analytical expression for 
$K(k,k')$: 
\begin{eqnarray}
K(k,k')&=& \frac{2\pi k'}{k}\int_{|k'-k|}^{k'+k}dq 
\frac{q}{1+\alpha q^2 +\beta q^4} \nonumber \\ 
&=& \frac{\pi k'}{k}\int_{(k'-k)^2}^{(k'+k)^2}\frac{dx}{1+\alpha x + \beta x^2} 
\nonumber \\ 
&=&\frac{\pi k'}{k}\eta\{{\rm arctanh}[\xi(k'-k)^2 + \lambda] \nonumber \\ 
&-& {\rm arctanh} [\xi(k'+k)^2 + \lambda]\},
\label{eq:appeq7}  
\end{eqnarray}
where $\eta=1.2055\cdot 10^{-5}$, $\xi = 2.5822 \cdot 10^3$, and 
$\lambda = 1.0307$.  

Numerical plots of $K(k,k')$ for various values of $k$ shows that it 
can be roughly approximated by a Gaussian,
\begin{equation} 
K(k,k') \propto 
\exp\left(-\frac{(k'-\mu_k)^2}{2\sigma_k^2}\right),
\label{eq:appeq8}
\end{equation} 
and that for $k > 0.1\;h\,{\rm Mpc}^{-1}$, 
$\sigma_k \ll k$, and $\mu_k \approx k$, 
so that the Gaussian approaches $\delta(k'-k)$.  In this regime of $k$, 
we will therefore have 
\begin{equation} 
\hat{P}_{\rm conv} (k)\propto \int_0^\infty \delta(k'-k)P_g(k')dk' 
=P_g(k).
\label{eq:appeq9}
\end{equation} 
For $k \ll 0.1\;h\,{\rm Mpc}^{-1}$, 
the Gaussian is very broad, $\sigma_k \gg k$ 
and $\sigma_k \gg \mu_k$. To illustrate what happens in this regime of 
$k$, we take  $P_g(k) \propto k$ for $k < k_b$ and 
$P_g(k)\propto k^{-2}$ for $k > k_b$.  Then 
\begin{eqnarray} 
\hat{P}_{\rm conv} (k)&\propto& \int_0 ^{k_b}
\exp\left(-\frac{(k'-\mu_k)^2}{2\sigma_k^2}\right)k'dk' \nonumber \\ 
&+& \int_{k_b} ^\infty \exp\left(-\frac{(k'-\mu_k)^2}{2\sigma_k^2}\right) 
\frac{dk'}{k'^2} . \label{eq:appeq10} 
\end{eqnarray} 
Both integrals can be evaluated analytically in terms of the error 
function ${\rm erf}(x)$.  Numerically, it turns out that it is a 
reasonable approximation in this regime to replace the Gaussian by its 
peak value, which gives 
\begin{eqnarray}
\hat{P}_{\rm conv} (k) &\propto& \int_0 ^{k_b}k' dk' + \int_{k_b} ^\infty \frac{dk'}
{k'^2} \nonumber \\ 
&=& \frac{k_b^2}{2} + \frac{1}{k_b} \approx \frac{1}{k_b}. \label{eq:appeq11} 
\end{eqnarray}
Note that this is independent of $k$, which agrees with the numerical 
results for the convolved $P(k)$.

\end{document}